\documentclass[sigconf, review]{acmart}

\setcopyright{none}
\authorsaddresses{}
\renewcommand\footnotetextcopyrightpermission[1]{}

\usepackage[normalem]{ulem}

\usepackage{caption}
\usepackage{enumitem}
\newcommand{\method}{NASiC}

\usepackage{etoolbox} 

\makeatletter
\renewcommand\@seccntformat[1]{%
  \csname the#1\endcsname
  \ifstrequal{#1}{section}{\hspace{0.6em}}{%
    \ifstrequal{#1}{subsection}{\hspace{0.4em}}{%
      \ifstrequal{#1}{subsubsection}{\hspace{0.35em}}{\quad}%
    }%
  }%
}
\makeatother

\makeatletter
\def\@secfont{\bfseries\Large}

\def\@subsecfont{\bfseries\large}
\def\@subsubsecfont{\bfseries\normalsize}

\makeatother

\begin{document}

\title{NASiC: 3D \uline{N}AND-based C\uline{A}M-\uline{S}elected Mult\uline{i}bit \uline{C}IM Architecture for Efficient On-Device Mixture-of-Experts LLM Inference}

\vspace{-5pt}

 \author{\fontsize{11}{13}\selectfont Weikai Xu\textsuperscript{1}, Meng Li\textsuperscript{2,1,3}*, Shuzhang Zhong\textsuperscript{2,1}, Tianyang Luo\textsuperscript{1}, Dongxue Zhao\textsuperscript{1}, Ling Liang\textsuperscript{1,3}*,\\ Zongwei Wang\textsuperscript{1,3}*, Qianqian Huang\textsuperscript{1,3}, Yimao Cai\textsuperscript{1,3}, Ru Huang\textsuperscript{1,3}*}
 \affiliation{%
   \institution{%
     \textsuperscript{1}School of Integrated Circuits, Peking University, Beijing, China; \textsuperscript{2}Institute for Artificial Intelligence, Peking University, Beijing, China; \textsuperscript{3}Beijing Advanced Innovation Center for Integrated Circuits, Beijing, China;\\
     *Corresponding author: ruhuang@pku.edu.cn; meng.li@pku.edu.cn; wangzongwei@pku.edu.cn; lingliang@pku.edu.cn.%
   }
   \country{}
 }

\begin{abstract}
The Mixture-of-Experts (MoE) models have emerged as the state-of-the-art paradigm for scaling up large language models (LLMs) without proportionally increased computational cost.
However, its on-device deployment faces a critical challenge due to the large memory requirement for storing all expert parameters.
3D NAND-based computing-in-memory (CIM) architectures uniquely offer high storage capacity and reduced data movement, while they are ill-suited for MoE models with dynamically sparse expert activation, leading to a degradation of effective computational parallelism, along with underutilization of multibit storage capability of Flash cells.
In this work, we proposed a 3D NAND-based content addressable-selected CIM architecture, dubbed as \method~, which is tailored to MoE models. 
By leveraging the intrinsic string structure of 3D NAND technology, \method~ fuses the dynamical expert selection through CAM-based masking mechanism and activated expert computation through CIM into a single computation cycle, eradicating redundant computation and enhancing computational parallelism.
Moreover, circuit-level optimizations and multibit CIM cell are co-designed with proposed \method~ architecture, featuring block-wise parallel computation with \textit{in-situ} signed multibit input and weight expansion, substantially improving the throughput and energy-efficiency of NAND CIM array, as well as the utilization of high-density 3D NAND technology for MoE models.
With extensive experimental results, we demonstrate \method~achieves 4$\sim$114.8$\times$ improved performance and 3.9$\sim$70$\times$ improved energy efficiency over state-of-the-art designs, along with high accuracy, showing its great potential for efficient on-device MoE LLM inference.
\end{abstract}

  
\pagestyle{plain}
\maketitle
\thispagestyle{plain}

\vspace{-1em}
\section{Introduction}
\begin{figure}[tb]
    \centering
    \includegraphics[width=1\linewidth]
    {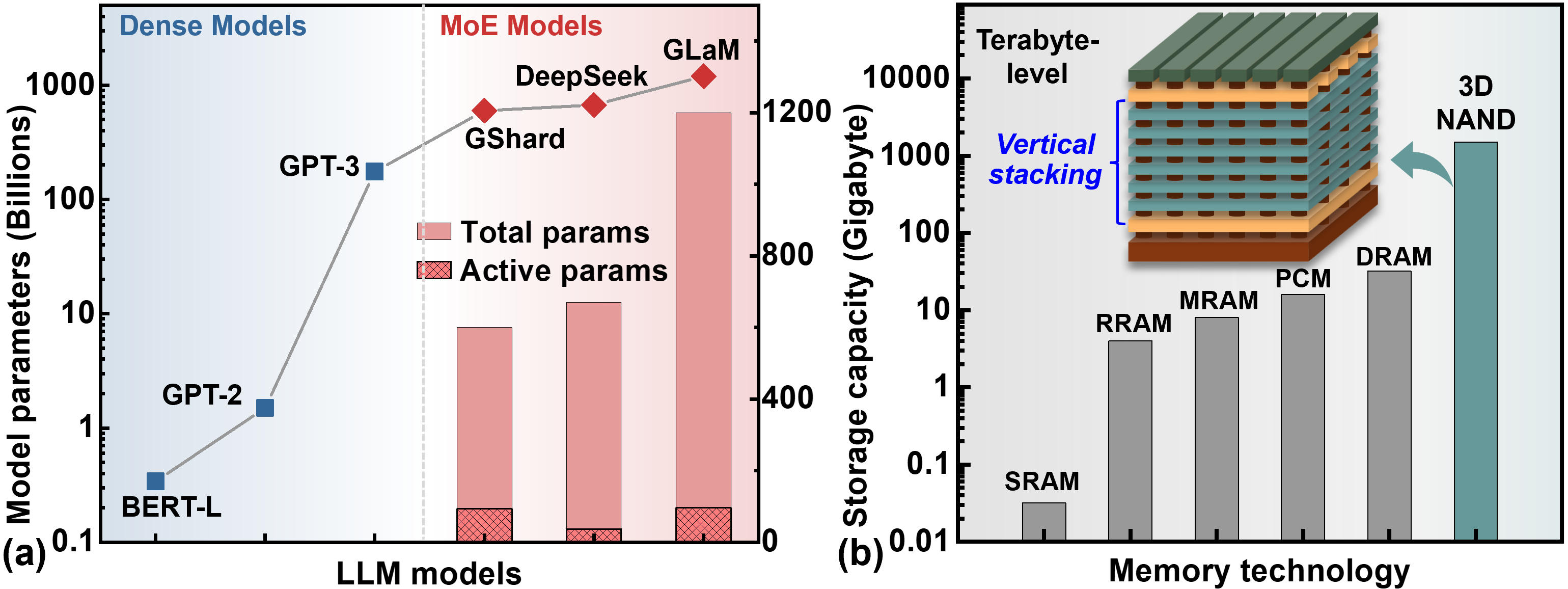}
    \vspace{-20pt}
    \caption{(a) The scaling trend of large language models (LLMs) from dense models to Mixture-of-Experts (MoE) models. (b) Storage capacity comparison of memory technologies, identifying 3D NAND as the unique terabyte-level solution.}
    \Description{}
    \vspace{-22pt}
    \label{fig:1}
\end{figure}

\vspace{-0.1em}
Transformer-based large language models (LLMs) have achieved remarkable success in various generative AI applications, such as dialogue system, document summarization, code generation, and multimodal reasoning \cite{vaswani2017attention, achiam2023gpt, touvron2023llama, yi2024survey,sudhakaran2024mariogpt,van2024adapted,li2025perception}.
The ever-increasing capabilities of LLMs are largely driven by the continuous scaling of model parameters to the trillion-level, while imposing significant challenges for computational efficiency \cite{kaplan2020scaling,chowdhery2023palm,guo2025survey}. 
To address this, the Mixture-of-Experts (MoE) model enables parameter scaling by increasing the number of expert parameters while activating them sparsely, thereby avoiding the proportional increase in computational cost associated with dense models and establishing it as a mainstream architecture in state-of-the-art LLMs
\cite{shazeer2017outrageously,lepikhin2020gshard,dai2024deepseekmoe,du2022glam, fedus2022switch, costa2022no} (Figure \ref{fig:1}a).
Although MoE model has proven to exhibit comparable or even-higher performance than dense models, its on-device deployment still faces a critical challenge of large memory requirement for storing all expert parameters \cite{zhao2025insights,yi2023edgemoe,chen2025collaborative}.
Among current memory technologies, 3D NAND uniquely offers the highest storage capacity, reaching the terabyte level by vertically stacking hundreds of layers with multi-level storage capacity per Flash cell, which is necessary for on-device MoE deployment \cite{yu2024semiconductor, yu2024cambricon, lee2025aif} (Figure \ref{fig:1}b).
Previous works based on von Neumann architecture offload MoE parameters to 3D NAND, which are then fetched on-demand into the GPU memory for computation, incurring substantial power and latency cost from the massive data movement \cite{huang2023towards,hwang2024pre,kyung2025ssd}.

Recently, some works have investigated compute-in-memory (CIM) architectures based on 3D NAND structure to reduce data movement, by implementing the general matrix-vector multiplication (GEMV) operations within the storage array \cite{hsu2020approach, lue2019optimal,shim2020technological,shim2020architectural}.
However, existing 3D NAND-based CIM designs primarily rarely exploiting the unique 3D NAND structure for co-optimization with MoE models, thereby still facing significant challenges for efficient on-device MoE inference.
\textbf{\emph{Challenge 1:}} Current 3D NAND CIM architectures are ill-suited for dynamically sparse MoE workloads due to the high-aspect-ratio rectangular layout of each 3D NAND CIM layer, which leads to redundant computation and a degradation of effective computational parallelism.
\textbf{\emph{Challenge 2:}} Current 3D NAND CIM circuits suffer a fundamental degradation of computational parallelism, constrained by the 3D NAND block structure and inefficient multibit input expansion strategies.
\textbf{\emph{Challenge 3:}} Current 3D NAND CIM technologies exclusively employ a single-level Flash cell, which underutilizes the intrinsic high-capacity benefit of 3D NAND technology, thereby making it difficult to store the MoE model parameters.
Therefore, it is still challenging for current 3D NAND-based CIM designs to efficiently support MoE models with both massive storage requirements and sparse computation workloads, restricting the practical deployment at the edge.

In this work, we proposed \method~, a 3D \uline{N}AND-based content addressable memory (C\uline{A}M)-\uline{S}elected mult\uline{i}bit \uline{C}IM architecture, tailored to the high-density storage and sparse computation requirements of MoE models.
\method~addresses the challenges mentioned above, and the main contributions can be summarized as follows:
\begin{itemize}[leftmargin=2em,labelsep=0.5em]
    \item \textbf{3D NAND-based CAM-selected CIM architecture:} By leveraging the intrinsic string structure of 3D NAND technology, we propose a novel CIM architecture incorporated CAM-based masking functionality for dynamically selecting the activated experts.
    Besides, an efficient hardware mapping strategy of MoE experts is further devised, which significantly alleviates the redundant computation with increased computational parallelism, thereby maximizing the effective utilization of the 3D NAND-based CIM architecture for sparse MoE computation.
    \item \textbf{Circuit-level optimizations for 3D NAND CIM array:} To enhance intrinsic computational parallelism of the 3D NAND CIM array, we propose a modified thermometer encoding scheme adapted for NAND block structure with \textit{in-situ} multibit input expansion, enabling block-wise parallel computation with improved computational parallelism of CIM array.
    \item \textbf{3D NAND-based multibit CIM cell:} By leveraging the multi-level storage capability of Flash cells, we propose a 3D NAND-based multibit CIM cell design and align it with block-wise encoding scheme, thereby fully exploiting the high-density 3D NAND for MoE models. 
    Moreover, we optimize the operational timing of multibit CIM based on 3D NAND structural properties, which can eliminate the redundant array operations and enhance the overall computational energy efficiency.
    \item {Experiments and evaluations show that the proposed \method~architecture with circuit- and cell-level co-optimization, can achieve 4$\sim$114.8$\times$ improved performance and 3.9$\sim$70$\times$ improved energy efficiency over state-of-the-art designs, showing its great potential for efficient on-device MoE inference.}
\end{itemize}

\vspace{-1em}
\section{Background}

\subsection{MoE-based sparse LLM model} 
\begin{figure}[tb]
    \centering
    \includegraphics[width=1\linewidth]
    {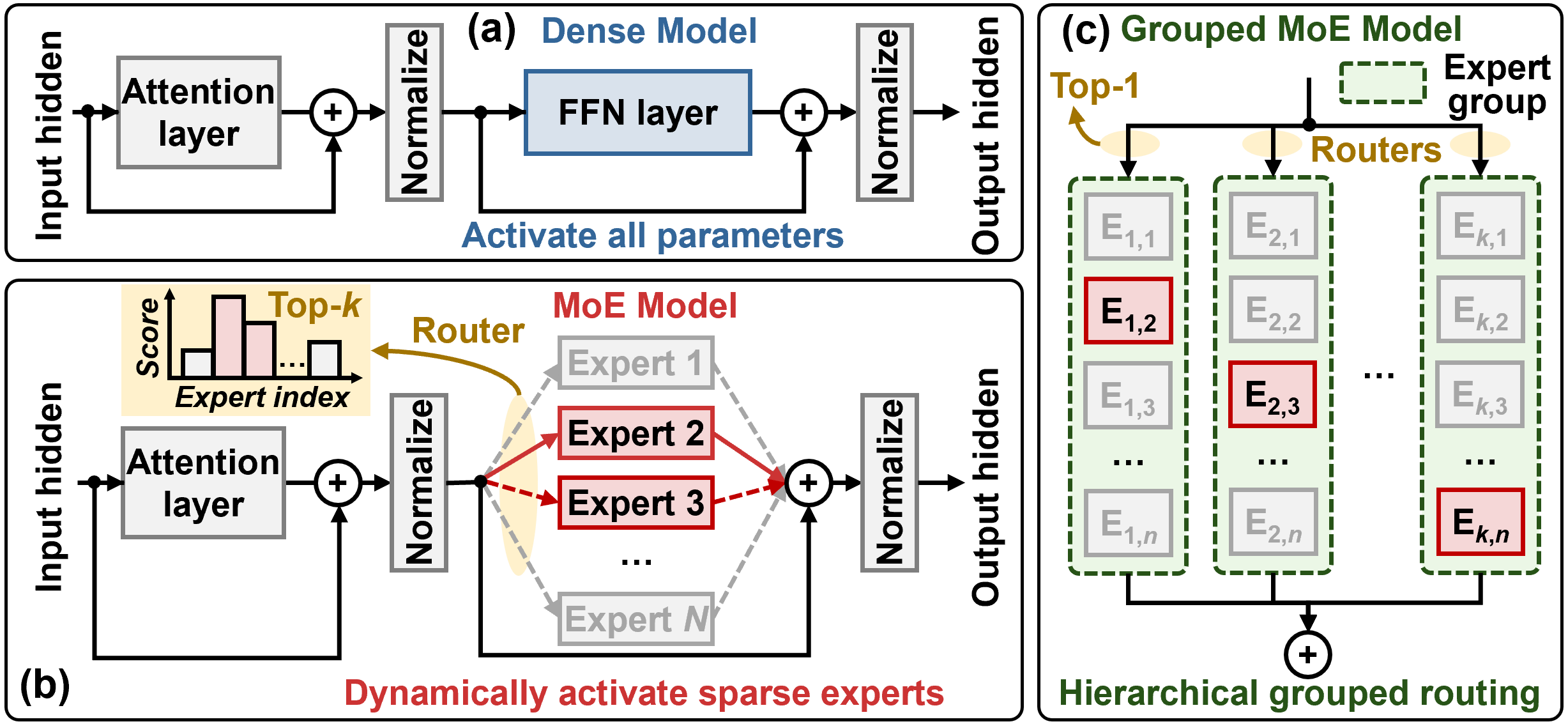}
    \vspace{-20pt}
    \caption{(a) Conventional dense model with the feed-forward network (FFN) layer. (b) The standard MoE model. (c) The grouped MoE model.}
    \Description{}
    \vspace{-10pt}
    \label{fig:2}
\end{figure}

\vspace{-0.2em}
The continuous scaling of LLM parameters has been the primary driver to enhanced performance, which leads to a proportional increase in computational cost, largely dominated by the feed-forward network (FFN) layers \cite{vaswani2017attention, achiam2023gpt, touvron2023llama, kaplan2020scaling}.
As shown in Figure \ref{fig:2}(a), the conventional dense model activates the entire FFN parameters for every input token, resulting in a significant computational bottleneck as the model size increases. 
The MoE model is designed to replace the single dense FFN layer with a set of \textit{N} parallel experts, and only a sparse subset of (i.e., top-\textit{k}) experts are dynamically activated by a router for each input token \cite{shazeer2017outrageously,lepikhin2020gshard,dai2024deepseekmoe} (Figure \ref{fig:2}b). 
MoE model effectively decouples the total parameter count from the computational cost, allowing it to scale to trillions of parameters while maintaining the computational overhead of a much smaller dense model \cite{du2022glam}.
Recently, some variants of MoE model have been proposed for efficient hardware mapping, such as the grouped MoE model \cite{tang2025pangu}, which organizes experts into distinct groups (e.g., \textit{k} groups) and selects one expert from each group (Figure \ref{fig:2}c). 
However, all MoE models share a fundamental challenge of massive storage requirement for storing all experts, which is the primary bottleneck for deployment on resource-constrained edge devices. 
Besides, state-of-the-art MoE models activate only small expert subset per token (e.g., top-1 in Switch Transformer \cite{fedus2022switch} and top-2 in NLLB-MoE \cite{costa2022no}), which results in a highly sparse computational workload, posing challenges for efficient hardware acceleration.

\begin{figure}[tb]
    \centering
    \includegraphics[width=1\linewidth]
    {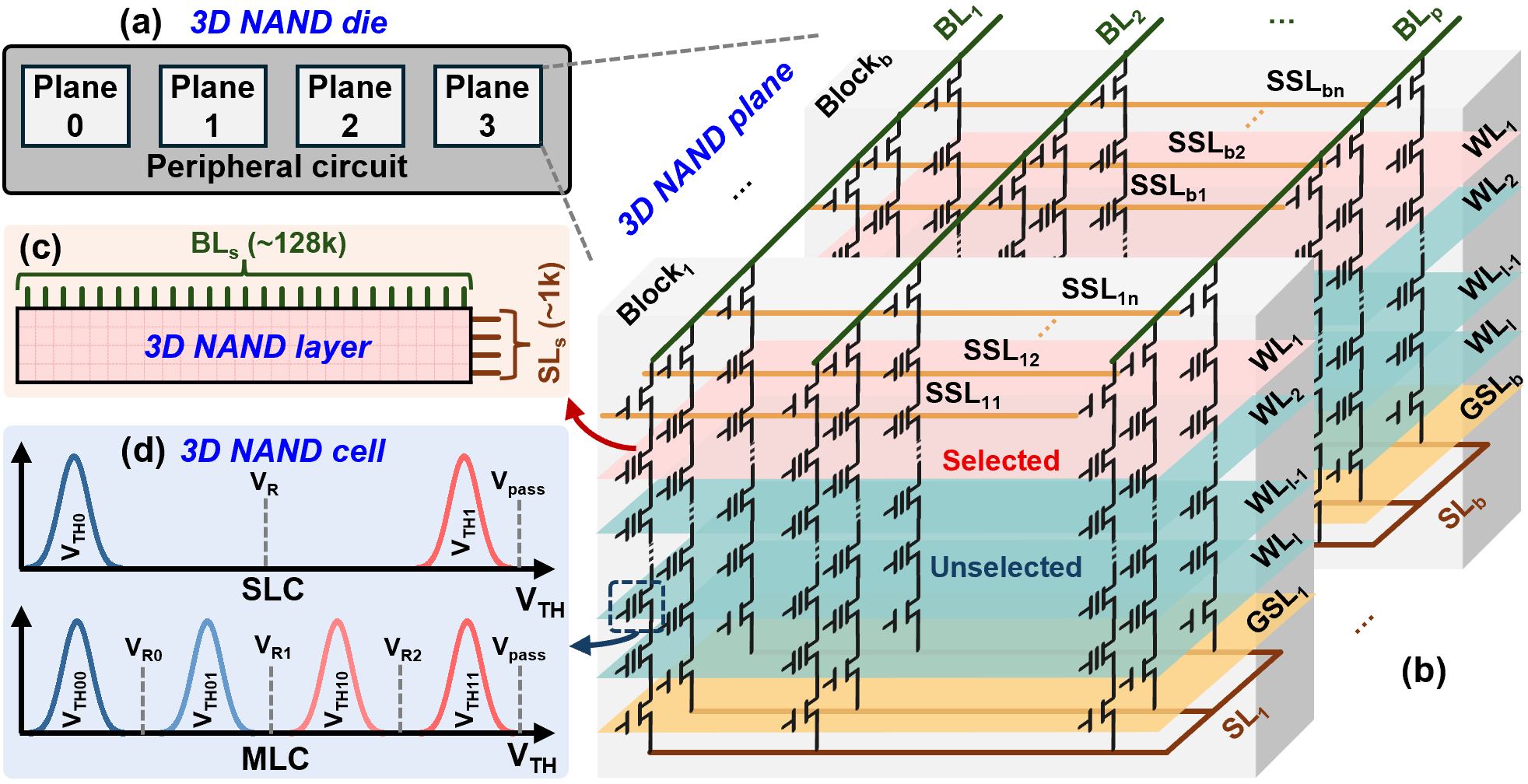}
    \vspace{-20pt}
    \caption{(a) Illustration of 3D NAND die. (b) Structure of 3D NAND plane. (c) Top-down view of 3D NAND layer. (d) Threshold voltage (V\textsubscript{TH}) distributions for Flash cell.}
    \Description{}
    \vspace{-15pt}
    \label{fig:3}
\end{figure}

\begin{figure*}[tb]
    \centering
    \includegraphics[width=1\linewidth]
    {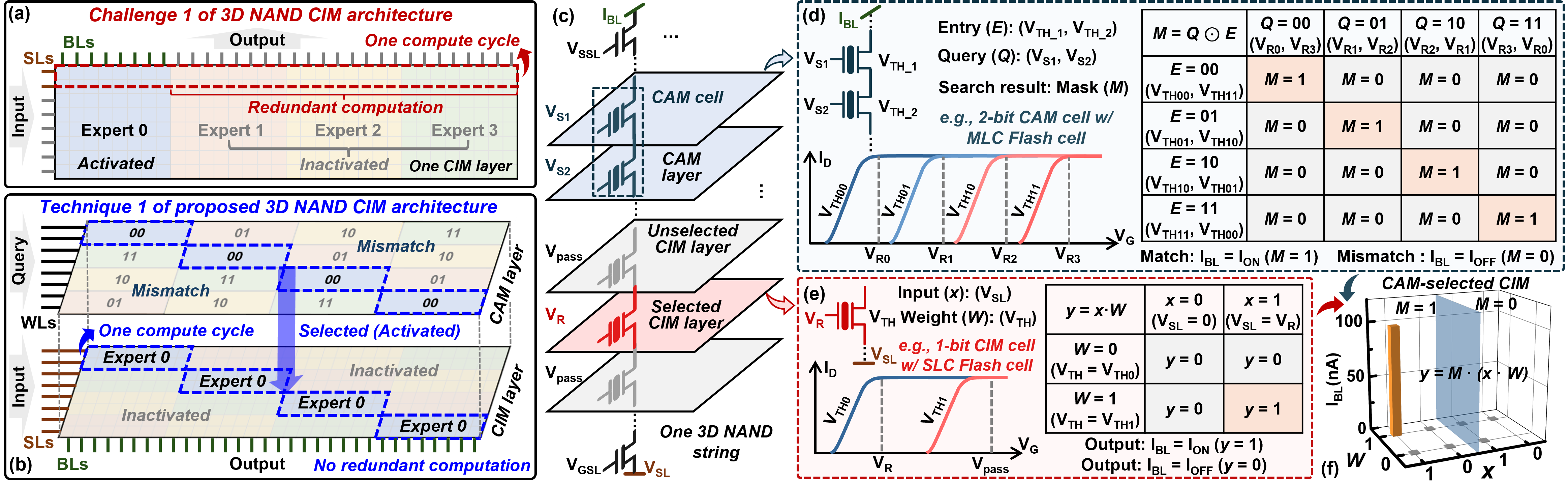}
    \vspace{-20pt}
    \caption{(a) Conventional 3D NAND CIM architecture. (b) The proposed 3D NAND-based content addressable memory (CAM)-selected CIM architecture. (c) The physical structure. (d) The 2-bit CAM cell design using multi-level cell (MLC) Flash cell. (e) The 1-bit CIM cell design. (f) Final computing results of the proposed CAM-selected CIM.}
    \Description{}
    \vspace{-16pt}
    \label{fig:4}
\end{figure*}

\vspace{-0.8em}
\subsection{3D NAND technology for CIM and MoE} 
\vspace{-0.2em}

As mentioned above, the massive storage requirement of MoE models necessitates a memory technology that offers unparalleled density. 
Among the memory technologies, 3D NAND is the only mature and commercially available technology that meets this demand, offering terabyte-level storage capacity achieved through 3D vertically stacking with multi-level storage capability \cite{kang201913,yuh20221}. 
Recently, 3D NAND-based CIM architectures have garnered significant attentions for LLM acceleration \cite{lue2019optimal,shim2020technological,shim2020architectural}. 
As shown in Fig. \ref{fig:3}ab, the 3D NAND die is composed of multiple planes, 
and one plane is composed of multiple blocks, each sharing a common ground select line (GSL) and source line (SL). 
Within each block, a set of vertically stacked Flash cells in series forms a string, which is connected to one bit line (BL) via a string select line (SSL) at the top, and to the common SL via the GSL at the bottom. 
Horizontally, all Flash cells on the same physical layer share a common word line (WL), where the cells selected by one SSL within a single block constitute a page. 
During computation, one layer of 3D NAND is activated at a time, and the GEMV operation is implemented (Fig. \ref{fig:3}c). 

Generally, two primary CIM designs are employed for 3D NAND, where inputs are applied to either BLs or SLs and the current is sensed on SLs or BLs, respectively. 
In 3D NAND structure, the number of BLs which defines the page size (e.g., 128k) is far greater than the number of SLs which is equal to the number of blocks (e.g., 1k) \cite{kang201913}. 
Besides, the input dimension is inherently constrained by device non-ideal effects, such as the device-to-device variation, and is thus even largely smaller than the number of SLs for practical applications.
Consequently, the SL-input scheme leverages the massive number of BLs as the output dimension can achieve higher effective computational parallelism for 3D NAND-based CIM \cite{shim2022impact}. 
Moreover, the Flash cell can be programmed to multiple non-overlapping threshold voltage (V\textsubscript{TH}) states 
from single-level cell (SLC) to multi-level cell (MLC) (Fig. \ref{fig:3}d), and even higher-density cells, making it a promising candidate for on-device MoE deployment \cite{yuh20221}. 

\vspace{-1em}
\section{Proposed \method~ Design for MoE}

\subsection{3D NAND-based CAM-selected CIM architecture}

\subsubsection{\textbf{Challenge 1: Inefficient 3D NAND CIM architecture for MoE}}
Conventional 3D NAND CIM architectures are primarily designed for dense computational workloads leveraging the high computational density \cite{hsu2020approach, lue2019optimal,shim2020technological,shim2020architectural}. 
However, they are inherently inefficient when processing the dynamically sparse workloads of MoE models, where the dense FFN layer is replaced by multiple parallel narrow experts, with relatively close dimensions of their weight matrices \cite{cai2025survey}.
Figure \ref{fig:4}(a) illustrates the hardware mapping for MoE based on conventional 3D NAND CIM architecture, using a simple top-1 routing example where one expert (e.g., expert 0) is activated from a set of four experts (i.e., expert 0$\sim$3).
Due to the high-aspect-ratio rectangular layout of each 3D NAND CIM layer, multiple experts need to be contiguously mapped and stored along the elongated page direction, and the input signals are broadcast to all parallel experts in the same layer. 
Consequently, the CIM operations are executed across all experts, which results in massive redundant computation stemming from the mismatch between the MoE workload and the hardware architecture.
Besides, as previously discussed, the practical input dimension of CIM is severely constrained by device non-ideal effects, forcing the computation for a single activated expert to be serialized over multiple cycles.
Therefore, the conventional 3D NAND CIM architecture is inefficient for dynamically sparse MoE workloads, severely degrading the effective computational parallelism and energy efficiency.

\vspace{-1em}
\subsubsection{\textbf{Technique 1 (T1): 3D NAND-based CAM-selected CIM architecture for sparse MoE workloads}}
To address the challenge of redundant computation of conventional 3D NAND CIM architecture, we propose a novel 3D NAND-based CAM-selected CIM architecture, \method, with interleaved expert mapping strategy as shown in Figure \ref{fig:4}(b).
The core design of \method~architecture is to fuse the expert selection through CAM and the expert computation through CIM into a single computation cycle, thereby eradicating redundant computation at the hardware level. 
By leveraging the special string characteristics of 3D NAND, \method~ architecture achieves the \textit{in-situ} integration of CAM cell and CIM cell on the same 3D NAND string, as depicted in Figure \ref{fig:4}(c).
The CAM layers located in the upper portion of 3D NAND strings are used to store the unique identifier (i.e., entry) of each expert , and the CIM layers in the lower portion store the expert's weight parameters.
When the MoE router selects an expert identifier (e.g., "00" for expert 0), the identifier is broadcast as a query signal the CAM layers of all blocks (Figure \ref{fig:4}b), thereby selecting the CIM portion associated with the activated expert for computation.

As shown in Figure \ref{fig:4}d, each CAM cell is constructed with two series-connected FG transistors on the 3D NAND string. The CAM function is designed to select and activate the corresponding CIM portion, which is essentially a logical operation. 
Therefore, we exploit the intrinsic current-limiting behavior of 3D NAND string and the multi-level storage capacity of Flash cell to design a compact and multibit CAM cell, to reduce the hardware overhead of CAM component.
As an example, we illustrate the 2-bit CAM cell design using MLC Flash cell. 
The 2-bit entry is stored as the respective V\textsubscript{TH} states of two FG transistors (V\textsubscript{TH1}, V\textsubscript{TH2}), and the 2-bit search query is correspondingly applied as the gate voltages (V\textsubscript{S1}, V\textsubscript{S2}) to these two transistors.
The search result, defined as mask ($M$), effectively performs a logical XNOR-like operation between the 2-bit entry and query, with the detailed encoding and matching schemes shown in the table of Fig. \ref{fig:4}(d).
Only when the search query and the stored entry are identical will both series-connected transistors turn on simultaneously, and the selected 3D NAND string conduct a high current, representing a match condition (i.e., $M=1$). 
All other combinations of query and entry will place the 3D NAND string in a non-conductive state, signaling a mismatch condition (i.e., $M=0$). We firstly implement a standard 1-bit CIM cell with SLC Flash cell on the same 3D NAND string (Figure \ref{fig:4}e), which perform the multiplication of the input ($x$) and the stored weight ($W$). 
The key innovation of proposed \method~ architecture is the natively gated computation mechanism illustrated in Figure \ref{fig:4}(f). By sharing the same 3D NAND string, the final output ($y$) current sensed on BL (I\textsubscript{BL}) is the logical AND between the select signal (i.e., $M$) of CAM and the computation result of CIM ($x \cdot W$), i.e., $y = M \cdot (x \cdot W)$.
This design ensures that only the CIM computation result ($x \cdot W$) of the expert which has been selected by the CAM layer ($M=1$) can contribute to I\textsubscript{BL}. 
Otherwise, for all mismatched experts ($M=0$), the computation is natively gated regardless of their computation results, which enables activating the input in parallel within a single computation cycle (Figure \ref{fig:4}b).
Consequently, \method~architecture eliminates the redundant computation and significantly improve the effective parallelism and energy efficiency for MoE inference.

\begin{figure}[tb]
    \centering
    \includegraphics[width=1\linewidth]
    {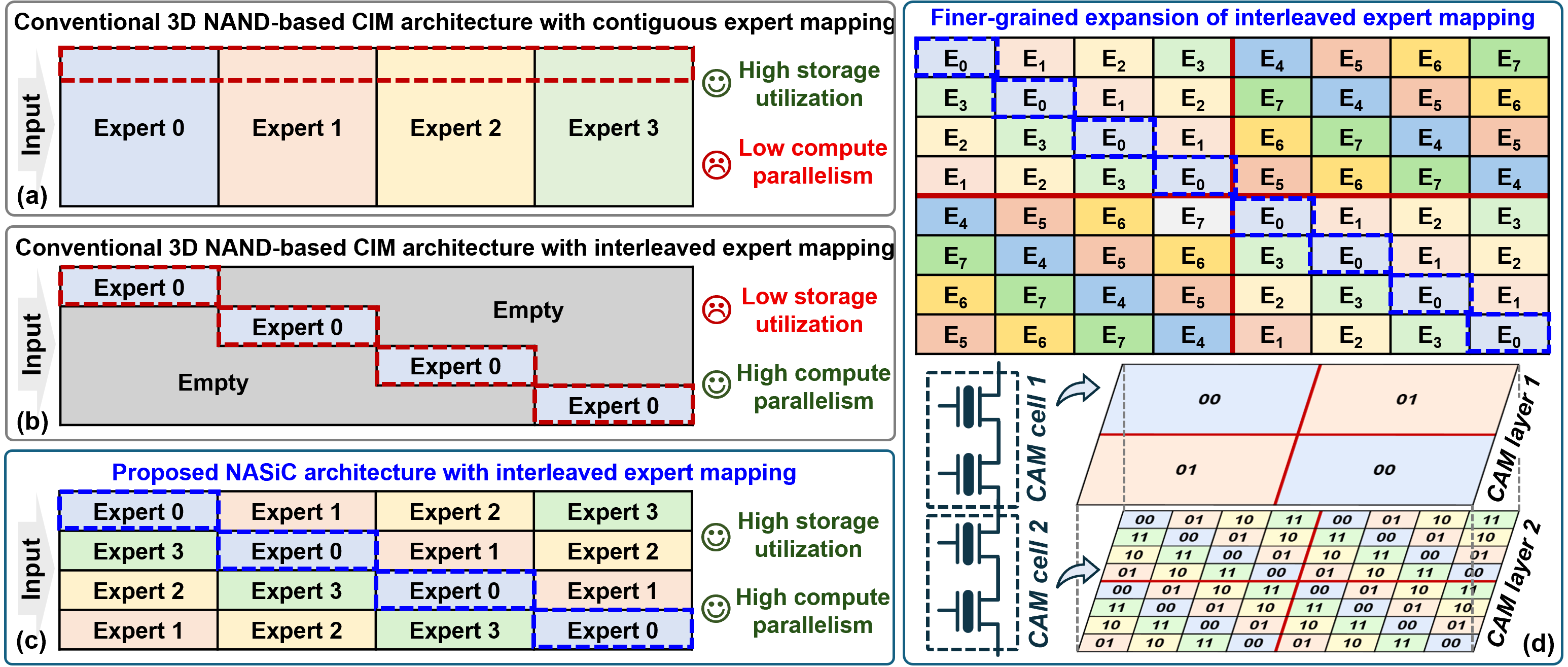}
    \vspace{-22pt}
    \caption{Conventional 3D NAND CIM architecture with (a) contiguous expert mapping strategy or (b) interleaved expert mapping strategy. (c) Proposed \method~architecture with interleaved expert mapping strategy. (d) Finer-grained expansion.}
    \Description{}
    \vspace{-20pt}
    \label{fig:5}
\end{figure}

It should be noted that adopting the interleaved expert mapping strategy based on the conventional 3D NAND CIM architecture, the high compute parallelism can also be achieved by distributing a single expert (e.g., expert 0), as shown in Figure \ref{fig:5}(ab).
However, without the effective \textit{in-situ} selection mechanism of CAM, this mapping strategy is forced to leave the remaining region empty to avoid the interference from other experts, resulting in extremely low storage utilization of 3D NAND.
The proposed \method~architecture combines the interleaved expert mapping strategy with CAM-selected mechanism, which simultaneously achieves high storage utilization and high compute parallelism (Figure \ref{fig:5}c).
Moreover, \method~architecture can further support finer-grained interleaved expert mapping, which can be achieved either by expanding the bit-width of CAM cell leveraging more V\textsubscript{TH} states of FG transistor, or by configuring multiple CAM cells in series on the same 3D NAND string.
Figure \ref{fig:5}(d) provides a concrete example where two CAM layers, configured as 1-bit and 2-bit CAM cells, are used to select one of eight experts, and a match condition is signaled only when the entries in both CAM layers are matched, thereby enabling a more flexible and finer-grained mapping of numerous MoE experts.

\vspace{-0.5em}
\subsection{Circuit-level optimizations for 3D NAND CIM}

\begin{figure}[tb]
    \centering
    \includegraphics[width=1\linewidth]
    {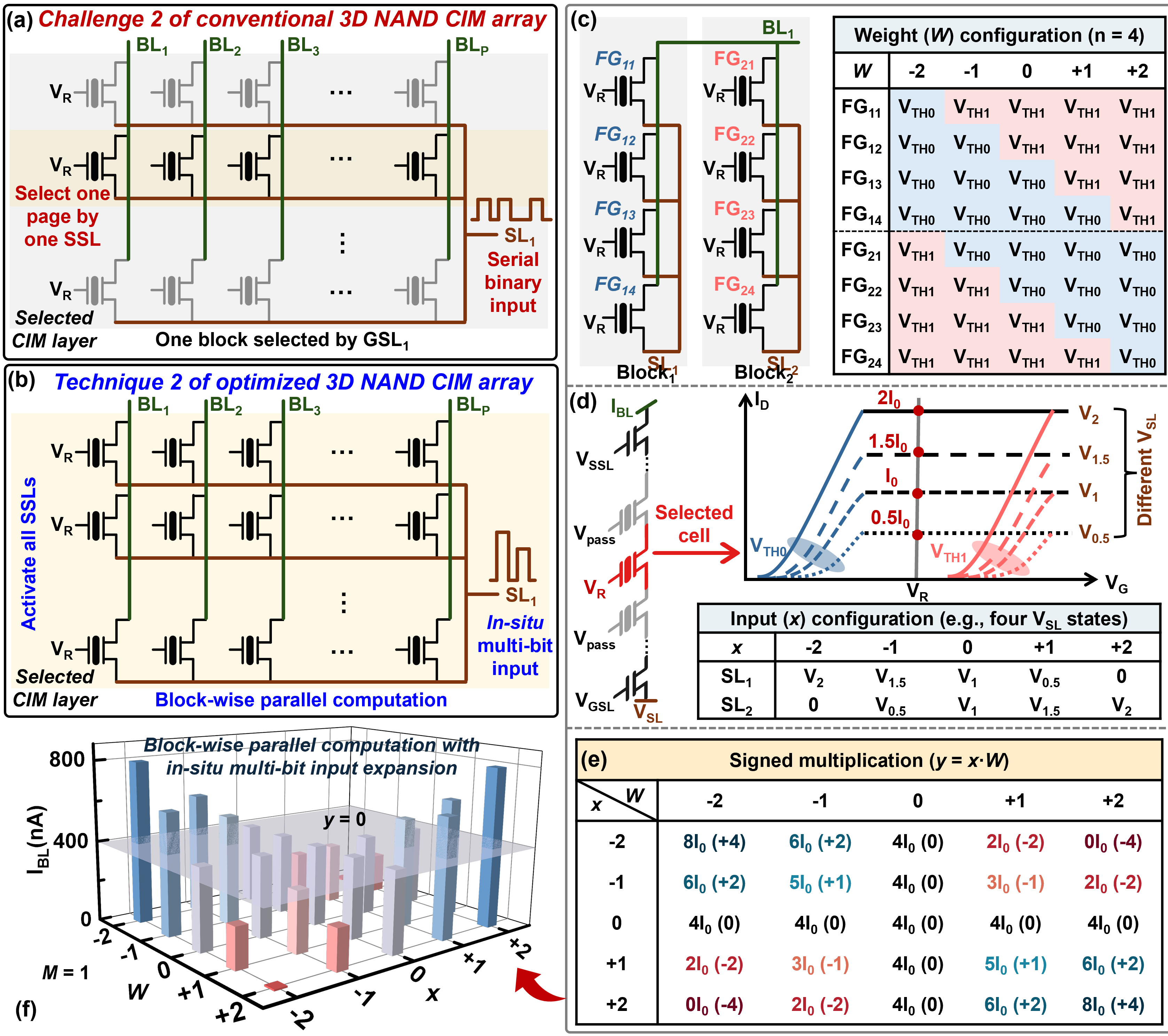}
    \vspace{-20pt}
    \caption{ (a) Conventional 3D NAND CIM circuit. (b) Proposed circuit-level optimizations. (c) The modified thermometer encoding scheme. (d) The \textit{in-situ} multibit input expansion scheme. (e) The signed multiplication map and (f) computation results.}
    \Description{}
    \vspace{-20pt}
    \label{fig:6}
\end{figure}

\subsubsection{\textbf{Challenge 2: Structurally constrained parallelism of conventional 3D NAND CIM circuits}}
Typically, in 3D NAND structure, each GSL corresponds to 4$\sim$8 SLLs for high integration density \cite{app11156703}, while this design severely limits computational parallelism, which can only select one page via SSL per block within a single computation cycle (Figure \ref{fig:6}a). 
Besides, conventional 3D NAND CIM designs generally employ serial binary input scheme, which applying computing signals bit-by-bit over time for multibit input expansion.
This approach is highly inefficient for MoE models, where input activations often require higher precision than weights, resulting in a degradation of effective throughput \cite{xiao2023smoothquant}.

\begin{figure*}[tb]
    \centering
    \includegraphics[width=1\linewidth]
    {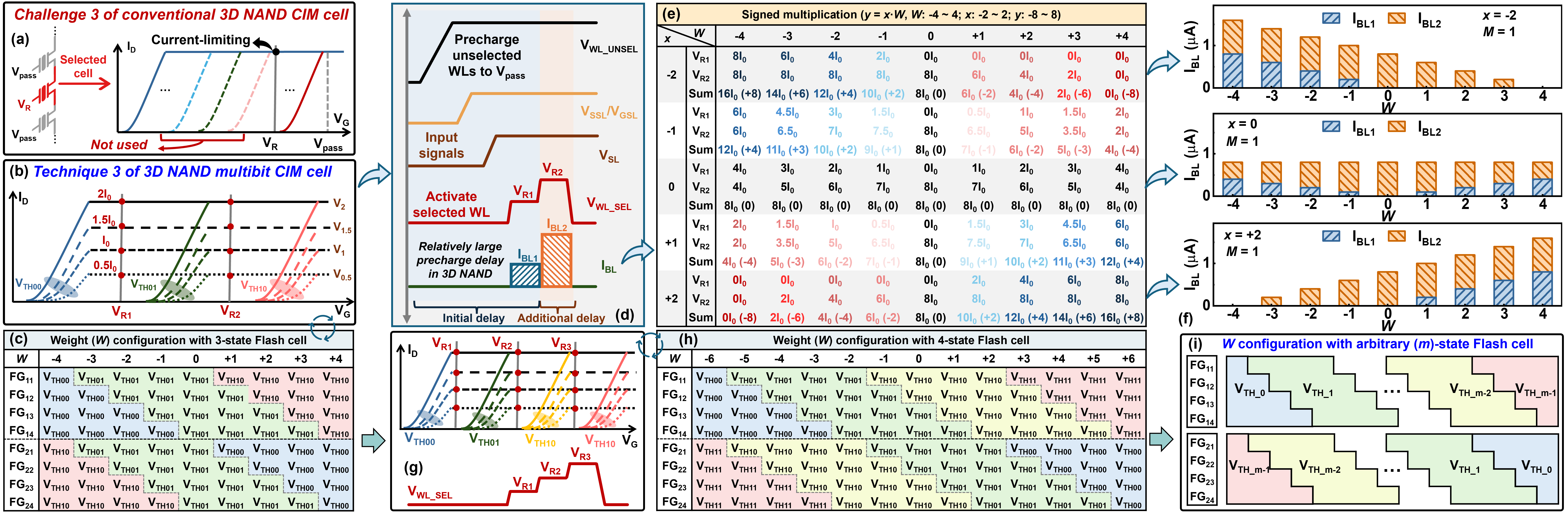}
    \vspace{-22pt}
    \caption{(a) Conventional 3D NAND CIM with SLC Flash. (b) Proposed 3D NAND-based multibit CIM cell. (c) Encoding scheme with 3-state Flash cell. (d) Optimized timing diagram of computation. (e) The signed multiplication map and (f) computing results. (g-h) Encoding scheme with 4-state Flash cell. (i) The weight expansion.}
    \Description{}
    \vspace{-15pt}
    \label{fig:7}
\end{figure*}

\subsubsection{\textbf{Technique 2 (T2): 3D NAND-based block-wise parallel computation with in-situ multibit input expansion}}
To overcome the structurally constrained parallelism, we propose circuit-level optimizations to enhance the intrinsic computational parallelism, predicated on the proposed CAM-selected CIM architecture.
To fully exploit the 3D NAND CIM array, we propose a modified thermometer encoding scheme adapted for 3D NAND block structure (Figure \ref{fig:6}b). 
Instead of treating multiple SLLs per block as independent pages, we activate them simultaneously to act as unified multibit weights for block-wise parallel computation.
We illustrate the encoding scheme using a standard 3D NAND plane where one GSL corresponds to four SSLs, where the magnitude of weight is encoded by the cumulative number of V\textsubscript{TH0} or V\textsubscript{TH1} states (Figure \ref{fig:6}c).
To further support signed weights, we adopt a dual-block configuration, where the FG transistors in the pair blocks are programmed to complementary 
V\textsubscript{TH} states.
In conjunction with the block-wise encoding scheme, we further propose develop an \textit{in-situ} signed multibit input expansion scheme (Figure \ref{fig:6}d).
By exploiting the current-limiting characteristics of 3D NAND strings, we modulate the voltage of SL (V\textsubscript{SL}) to induce proportional I\textsubscript{BL}, which represents the multi-level input states.
Taking four V\textsubscript{SL} states as an example, we co-design the input encoding scheme with block-wise weight encoding, thereby achieving signed multiplication between block-wise signed multibit weight and \textit{in-situ} signed multibit input (Figure \ref{fig:6}e). 
Figure \ref{fig:6}(f) presents the computing results, demonstrating high linearity and the enhanced computation parallelism.

It is worth emphasizing that while the proposed circuit-level optimizations are technically applicable to conventional 3D NAND CIM architecture, applying them in isolation yields negligible benefits, due to the inherently constrained input parallelism.
Within the proposed CAM-selected CIM architecture, these circuit-level optimizations directly translate to enhanced CIM array utilization.
This architecture-circuit co-design thereby maximizes the theoretical computational parallelism of 3D NAND CIM array.

\subsection{3D NAND-based multibit CIM cell}
\vspace{-0.3em}
\subsubsection{\textbf{Challenge 3: Storage underutilization of 3D NAND technology}}
As shown in Figure \ref{fig:7}(a), conventional 3D NAND CIM designs exclusively employ SLC storage of Flash cells, using only two V\textsubscript{TH} states of FG transistor, due to the current-limiting in 3D NAND strings.
This approach significantly underutilizes the intrinsic high-density multi-level Flash cells.
Consequently, it is challenging to store the massive parameters of MoE models, defeating the primary motivation for using 3D NAND technology.

\vspace{-0.8em}
\subsubsection{\textbf{Technique 3 (T3): 3D NAND-based multibit CIM cell with enhanced storage efficiency}}
To address the storage underutilization challenge, we propose a 3D NAND-based multibit CIM cell design, which fully leverages the storage density of 3D NAND technology.
Firstly, we co-design our above block-wise modified thermometer encoding scheme to compatible with the multi-level storage capability of NAND flash cells.
Taking a three-state Flash cell as an example (Figure \ref{fig:7}b), we introduce a scalable thermometer-based weight mapping, achieved by inserting additional V\textsubscript{TH} states to the basic SLC-based encoding scheme from Figure \ref{fig:6}(c), thereby upgrading the CIM cell from binary to multibit (Figure \ref{fig:7}c).
Specially, by adding one V\textsubscript{TH} state, the representable signed weight states are expanded from "-2"$\sim$"+2" to "-4"$\sim$"+4", significantly boosting storage density.
Moreover, to efficiently compute with the multibit CIM cell, we optimize the computation scheme, which involves multiple distinct read voltages (e.g., V\textsubscript{R1} and V\textsubscript{R2} for 3-state CIM cell) to the selected WL within a single computation cycle (Figure \ref{fig:7}d).
The final computation result is thus the summation of the I\textsubscript{BL} obtained from these multiple read voltage pulses, as detailed by the signed multiplication map in Figure \ref{fig:7}(e).
The computing results for each $x \cdot W$ combination showing distinct and linear current steps, demonstrating the feasibility of proposed high-density multibit CIM cell (Figure \ref{fig:7}f).
Furthermore, the proposed multibit CIM cell with the aligned block-wise encoding scheme has inherent scalability, which can be easily extended to 4-state CIM cell (Figure \ref{fig:7}gh) and arbitrary m-state CIM cell (Figure \ref{fig:7}i) with structural mapping. 
This demonstrates the flexibility and generality of our design, enabling 3D NAND CIM with even higher storage densities for MoE.

Notably, the additional delay introduced by our scheme is minimal and accounts for only a small fraction of the total cycle time, due to the relatively large parasitic RC time (e.g., unselected WLs precharging) in 3D NAND structure (Figure \ref{fig:7}d).
Therefore, the proposed 3D NAND-based multibit CIM cell with optimized computing scheme achieves a significant boost in storage density at a negligible delay cost.
In fact, it enables higher data parallelism, ultimately enhancing the overall performance of 3D NAND CIM.

{\renewcommand{\thefigure}{\Roman{figure}}
\begin{figure}[tb]
    \vspace{-10pt}
    \renewcommand{\figurename}{Table} 
    \setcounter{figure}{0} 
    \captionsetup{justification=centering}
    \caption{Configuration of 3D NAND plane.} 
    \vspace{0pt}
    \centering
    \includegraphics[width=1\linewidth]
    {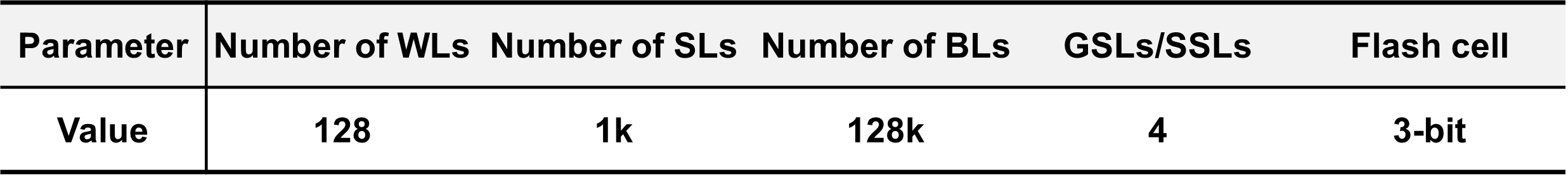}
    \vspace{-30pt}
    \label{table:1}
\end{figure}
}

\vspace{-1em}
\section{Experimental Results}

\vspace{-0.2em}
\subsection{Experiment setup}
We validate and evaluate the effectiveness of \method~architecture for on-device MoE inference at circuit and system levels.
The configuration of 3D NAND plane is based on a typical 3D NAND technology \cite{kang201913}, as detailed in Table \ref{table:1}.
For the circuit-level evaluations, we conducted SPICE simulations using a compact Verilog-A model of FG transistors \cite{chowdhury2024analysis} based on 65nm technology node, and considering the 3D NAND structural parasitic parameters derived from \cite{lee20223d}.
Moreover, we perform system-level analysis incorporating the impact of device non-idealities, and evaluate the key performance metrics including throughput, energy efficiency and area efficiency.

\vspace{-1em}
\subsection{Main results}
\vspace{-0.2em}
\subsubsection{Throughput} To ensure a realistic assessment, we first investigate the impact of device non-idealities on performance of MoE models.
As the standard deviation ($\sigma$) of current in Flash cells increases, the allowed input dimension is decreased to prevent accuracy dimension (Figure \ref{fig:8}a).
This constraint critically impacts the throughput of the conventional 3D NAND CIM architecture (Base) for MoE, where higher $\sigma$ requires more serialized computation cycles, each of which suffers from massive redundant computation.
Benefiting from the CAM-selection mechanism, our proposed \method~architecture with T1 eliminates this redundant computation despite the current variation, thereby dramatically boosting effective throughput (Figure \ref{fig:8}b).
Moreover, with finer-grained interleaved expert mapping, the throughput improvement becomes even more pronounced as $\sigma$ increases, growing from 4$\times$ to 16$\times$ over the Base.
We then perform an ablation study to evaluate the cumulative benefits of \method~architecture with three techniques, using an experimentally-derived  
$\sigma$ of 0.15 \cite{lue2019optimal} (Figure \ref{fig:8}c).
The throughput is further improved by applying T2 with enhanced intrinsic computational parallelism of the CIM array, and the improvement is more significant as the activated expert ratio ($r$) becomes sparser, which is preferable for highly sparse MoE models.
Besides, the throughput enhancement from T3 is agnostic to $r$, due to the applicable application involving multiple distinct read pulses and sensing phases within one computation cycle.
As we increase the states of the CIM cell, the number of operations scales linearly, while the total delay composed of additional read delay (t\textsubscript{2}) and fixed initial delay (t\textsubscript{1}) (Figure \ref{fig:7}d) increases sub-linearly, thereby resulting in a dramatic boost to effective computational throughput (Figure \ref{fig:8}d).

\begin{figure}[tb]
    \centering
    \setcounter{figure}{7} 
    \includegraphics[width=0.98\linewidth]
    {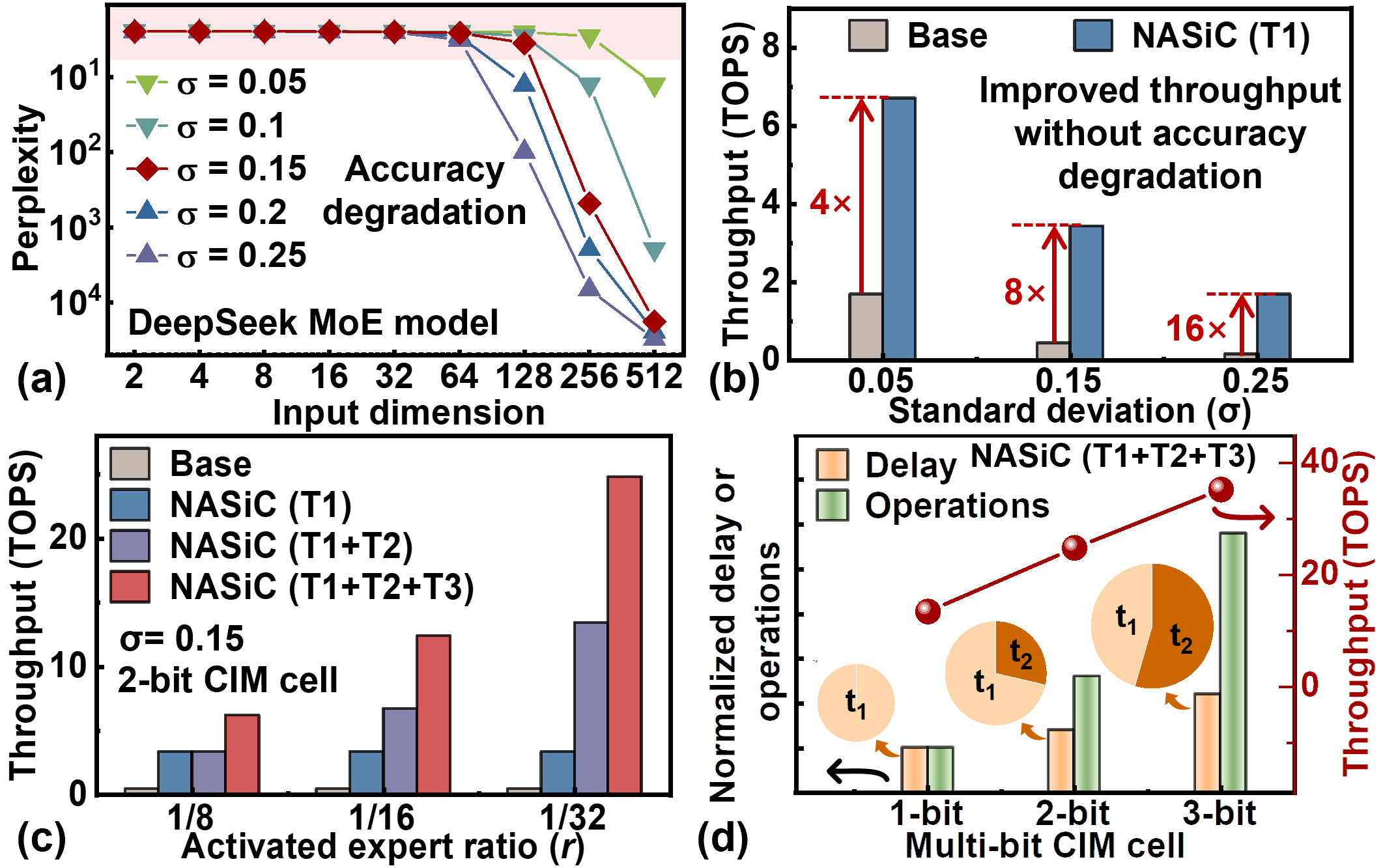}
    \vspace{-12pt}
    \caption{(a) Impact of device variation on perplexity. (b) Throughput gain of \method~architecture with T1 over the conventional architecture (Base). (c) Ablation study of \method~architecture with T1$\sim$T3. (d) Evaluation of the multibit CIM design in T3.}
    \Description{}
    \vspace{-13pt}
    \label{fig:8}
\end{figure}

\begin{figure}[tb]
    \centering
    \setcounter{figure}{8} 
    \includegraphics[width=0.95\linewidth]
    {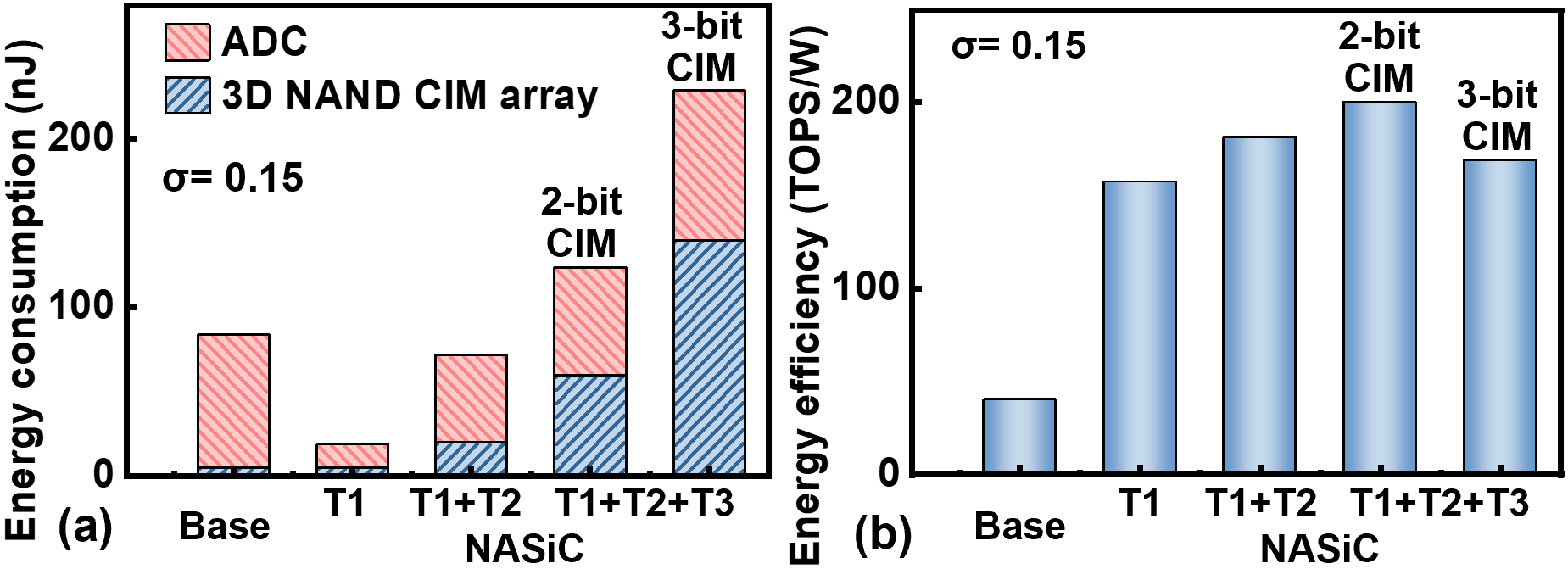}
    \vspace{-12pt}
    \caption{(a) Energy consumption breakdown of Base and \method~architecture with three techniques. (b) Energy efficiency gain of \method~architecture over Base.}
    \Description{}
    \vspace{-18pt}
    \label{fig:9}
\end{figure}

\vspace{-3pt}
\subsubsection{Energy efficiency}
The evaluations of energy consumption and efficiency are carried out with 128 input dimension ($\sigma$ = 0.15), and thus a 8-bit SAR ADC is used for quantization \cite{miki20172}.
Figure \ref{fig:9}(a) presents the energy consumption breakdown, which primarily consists of the 3D NAND CIM array and the ADC. 
Compared with the Base, our proposed \method~architecture with T1 significantly reduces total energy consumption by eliminating redundant computation.
In this scenario, the energy consumption is dominated by the 3D NAND array, largely due to the high precharging energy of 3D NAND signal lines.
As we apply T2 and T3, the computational density of the 3D NAND CIM array increases, leading to a corresponding rise in total energy, where the energy consumption of the ADC progressively overtakes that of the 3D NAND CIM array.
As shown in Figure \ref{fig:9}(b), our proposed \method~architecture with three techniques improve the energy efficiency from 3.9$\times$ to 5.1$\times$ compared with the Base.
Furthermore, when increasing the CIM cell to 3-bit with higher throughput (Figure \ref{fig:8}d), the energy efficiency is decreased compared with 2-bit CIM, which is primarily due to the rapidly increasing ADC energy and latency overhead.
While it illustrates a fundamental performance-efficiency trade-off in practical hardware designs, our \method~architecture enables flexibly optimized configuration for speed or energy sensitive applications.

\begin{figure}[tb]
    \centering
    \setcounter{figure}{9} 
    \includegraphics[width=0.98\linewidth]
    {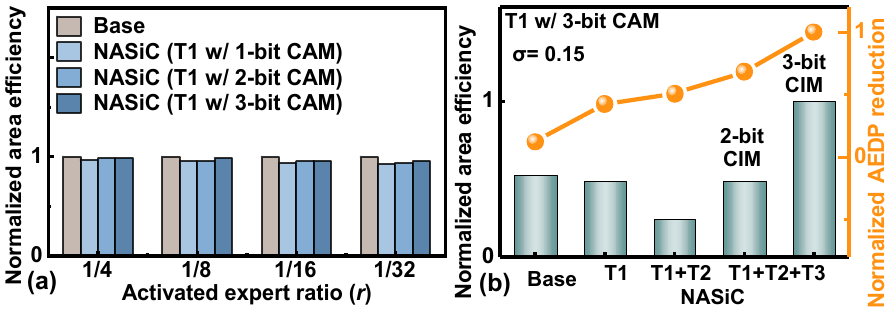}
    \vspace{-12pt}
    \caption{ (a) Impact of CAM layers in T1  for area efficiency. (b) Evaluation of area efficiency and overall area-energy-delay product of \method~architecture with T1$\sim$T3.}
    \Description{}
    \vspace{-18pt}
    \label{fig:10}
\end{figure}

{\renewcommand{\thefigure}{\Roman{figure}}
\begin{figure}[tb]
    \vspace{-7pt}
    \renewcommand{\figurename}{Table} 
    \setcounter{figure}{1} 
    \captionsetup{justification=centering}
    \caption{Comparison of proposed \method~architecture with other designs.} 
    \vspace{0pt}
    \centering
    \includegraphics[width=1\linewidth]
    {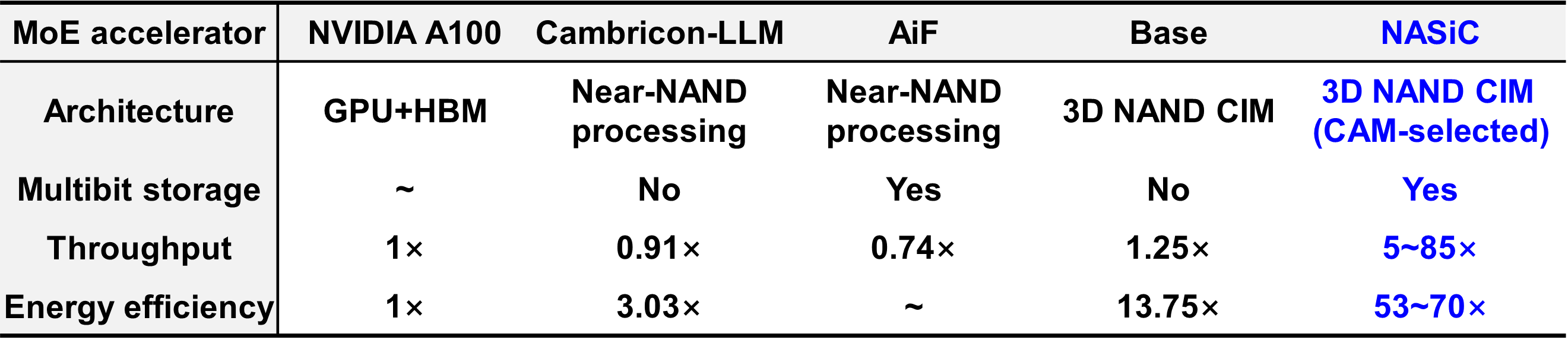}
    \vspace{-32pt}
    \label{table:2}
\end{figure}
}

\vspace{-0.5em}
\subsubsection{Area efficiency}
Since \method~architecture with T1 allocates a portion of 3D NAND layers for CAM-based expert selection, a slight reduction in storage utilization ( i.e., area efficiency) is expected (Figure \ref{fig:10}a).
Furthermore, we can implement the expert selection with fewer CAM layers by leveraging multibit CAM cells, resulting in a negligible reduction in area efficiency (< 0.05) even under highly sparse conditions ($r=1/32$) with 3-bit CAM configuration, which demonstrates effective expert activation with minimal area overhead.
As shown in Figure \ref{fig:10}(c), applying T2 results in a noticeable decrease in area efficiency, due to the trade-off between storage density and computational parallelism of modified thermometer encoding scheme.
However, this degradation is effectively reversed and surpassed by T3 with multibit CIM design, which further boost the area efficiency, utilizing the intrinsic high storage density of the 3D NAND technology.
Moreover, the overall comprehensive area-energy-delay product (AEDP) is consistently reduced by the \method~architecture with T1$\sim$T3, achieving a 3.5$\times$ to 8.3$\times$ reduction compared to the Base.

\vspace{-0.8em}
\subsection{Comparison with state-of-the-art designs}
\vspace{-0.3em}
We benchmark the proposed \method~architecture against state-of-the-art accelerators, including the NVIDIA A100 GPU\cite{choquette2021nvidia}, near-3D NAND processing architectures (Cambricon-LLM \cite{yu2024cambricon} and AiF \cite{ lee2025aif}), and the Base for MoE inference.
The proposed \method~architecture with cross-layer co-optimization achieves 5$\times$/5.5$\times$/6.8$\times$/4$\times$ to 85$\times$/93.4$\times$/114.8$\times$/68$\times$ improved throughput compared with GPU/Cambricon-LLM/AiF/Base, enabled by the progressive integration of the three proposed techniques.
Moreover, \method~architecture also achieves  53$\times$/17.5$\times$/3.9$\times$ to 70$\times$/23$\times$/5.1$\times$ improved energy efficiency compared with GPU/Cambricon-LLM/Base, indicating its great potential for on-device MoE inference.

\vspace{-1em}
\section{Conclusion}
\vspace{-0.3em}
In this work, a novel \method~architecture featuring integrated expert activation and computation, block-wise parallel computation with \textit{in-situ} signed multibit input and weight expansion, is proposed for dynamically sparse MoE workloads.
The \method~architecture eliminates the redundant computation and fully exploit the high-density multi-level storage capacity of Flash cells.
Evaluation results show significantly improved throughput and energy efficiency, indicating its great potential for on-device MoE inference.
\vspace{0.3em}

\noindent\textbf{Acknowledgments}: This work was supported in part by NSFC under Grant 62495102, Grant 92464104, and Grant 62341407, in part by the National Key Research and Development Program under Grant 2024YFB4505004, in part by Beijing Municipal Science and Technology Program under Grant Z241100004224015, in part by 111 Project under Grant B18001.

\bibliographystyle{unsrtnat}   

\bibliography{sample-base}

\end{document}